\documentclass[twocolumn,showpacs,preprintnumbers,superscriptaddress,amsmath,amssymb,PRL]{revtex4}
\usepackage{graphicx}
\usepackage{dcolumn}
\usepackage{float}
\usepackage{mathptmx, courier, pifont}
\usepackage[scaled=0.92]{helvet}
\usepackage[T1]{fontenc}
\usepackage{textcomp}
\usepackage{color}
\usepackage{booktabs}

\usepackage[dvipsnames]{xcolor}
\usepackage[colorlinks=true,urlcolor=Blue,linkcolor=Blue]{hyperref}
\usepackage[all]{hypcap}

\begin{document}

\title{Improving the resolving power of Isochronous Mass Spectrometry by employing an in-ring mechanical slit}

\author{J .H. Liu}
\affiliation{Key Laboratory of High Precision Nuclear Spectroscopy and Center for Nuclear Matter Science, Institute of Modern Physics, Chinese Academy of Sciences, Lanzhou 730000,  China}
\affiliation{School of Nuclear Science and Technology, University of Chinese Academy of Sciences, Beijing 100049, China}
\author{X.~Xu}\thanks{Corresponding author: xuxing@impcas.ac.cn}
\affiliation{Key Laboratory of High Precision Nuclear Spectroscopy and Center for Nuclear Matter Science, Institute of Modern Physics, Chinese Academy of Sciences, Lanzhou 730000, China}
\author{P.~Zhang}
\affiliation{Key Laboratory of High Precision Nuclear Spectroscopy and Center for Nuclear Matter Science, Institute of Modern Physics, Chinese Academy of Sciences, Lanzhou 730000,  China}
\affiliation{School of Nuclear Science and Technology, University of Chinese Academy of Sciences, Beijing 100049, China}
\author{P.~Shuai}
\affiliation{Key Laboratory of High Precision Nuclear Spectroscopy and Center for Nuclear Matter Science, Institute of Modern Physics, Chinese Academy of Sciences, Lanzhou 730000,  China}
\author{X.~L.~Yan}
\affiliation{Key Laboratory of High Precision Nuclear Spectroscopy and Center for Nuclear Matter Science, Institute of Modern Physics, Chinese Academy of Sciences, Lanzhou 730000,  China}
\author{Y.~H.~Zhang}
\affiliation{Key Laboratory of High Precision Nuclear Spectroscopy and Center for Nuclear Matter Science, Institute of Modern Physics, Chinese Academy of Sciences, Lanzhou 730000,  China}
\affiliation{School of Nuclear Science and Technology, University of Chinese Academy of Sciences, Beijing 100049, China}
\author{M.~Wang}\thanks{Corresponding author:  wangm@impcas.ac.cn} 
\affiliation{Key Laboratory of High Precision Nuclear Spectroscopy and Center for Nuclear Matter Science, Institute of Modern Physics, Chinese Academy of Sciences, Lanzhou 730000, China}
\affiliation{School of Nuclear Science and Technology, University of Chinese Academy of Sciences, Beijing 100049, China}
\author{Yu.~A.~Litvinov}
\affiliation{Key Laboratory of High Precision Nuclear Spectroscopy and Center for Nuclear Matter Science, Institute of Modern Physics, Chinese Academy of Sciences, Lanzhou 730000,  China}
\affiliation{GSI Helmholtzzentrum f{\"u}r Schwerionenforschung, Planckstra{\ss}e 1, Darmstadt, 64291 Germany}
\affiliation{Max-Planck-Institut f\"{u}r Kernphysik, Saupfercheckweg 1, 69117 Heidelberg, Germany}
\author{H.~S.~Xu}
\affiliation{Key Laboratory of High Precision Nuclear Spectroscopy and Center for Nuclear Matter Science, Institute of Modern Physics, Chinese Academy of Sciences, Lanzhou 730000,  China}
\affiliation{School of Nuclear Science and Technology, University of Chinese Academy of Sciences, Beijing 100049, China}
\author{K.~Blaum}
\affiliation{Max-Planck-Institut f\"{u}r Kernphysik, Saupfercheckweg 1, 69117 Heidelberg, Germany}
\author{T.~Bao}
\affiliation{Key Laboratory of High Precision Nuclear Spectroscopy and Center for Nuclear Matter Science, Institute of Modern Physics, Chinese Academy of Sciences, Lanzhou 730000,  China}
\author{H.~Chen}
\affiliation{Key Laboratory of High Precision Nuclear Spectroscopy and Center for Nuclear Matter Science, Institute of Modern Physics, Chinese Academy of Sciences, Lanzhou 730000,  China}
\affiliation{School of Nuclear Science and Technology, University of Chinese Academy of Sciences, Beijing 100049, China}
\author{X.~C.~Chen}
\affiliation{Key Laboratory of High Precision Nuclear Spectroscopy and Center for Nuclear Matter Science, Institute of Modern Physics, Chinese Academy of Sciences, Lanzhou 730000,  China}
\author{R.~J.~Chen}
\affiliation{Key Laboratory of High Precision Nuclear Spectroscopy and Center for Nuclear Matter Science, Institute of Modern Physics, Chinese Academy of Sciences, Lanzhou 730000,  China}
\author{C.~Y.~Fu}
\affiliation{Key Laboratory of High Precision Nuclear Spectroscopy and Center for Nuclear Matter Science, Institute of Modern Physics, Chinese Academy of Sciences, Lanzhou 730000,  China}
\author{D.~W.~Liu}
\affiliation{Key Laboratory of High Precision Nuclear Spectroscopy and Center for Nuclear Matter Science, Institute of Modern Physics, Chinese Academy of Sciences, Lanzhou 730000,  China}
\affiliation{School of Nuclear Science and Technology, University of Chinese Academy of Sciences, Beijing 100049, China}
\author{W.~W.~Ge}
\affiliation{Key Laboratory of High Precision Nuclear Spectroscopy and Center for Nuclear Matter Science, Institute of Modern Physics, Chinese Academy of Sciences, Lanzhou 730000,  China}
\author{R.~S.~Mao}
\affiliation{Key Laboratory of High Precision Nuclear Spectroscopy and Center for Nuclear Matter Science, Institute of Modern Physics, Chinese Academy of Sciences, Lanzhou 730000,  China}
\author{X.~W.~Ma}
\affiliation{Key Laboratory of High Precision Nuclear Spectroscopy and Center for Nuclear Matter Science, Institute of Modern Physics, Chinese Academy of Sciences, Lanzhou 730000,  China}
\author{M.~Z.~Sun}
\affiliation{Key Laboratory of High Precision Nuclear Spectroscopy and Center for Nuclear Matter Science, Institute of Modern Physics, Chinese Academy of Sciences, Lanzhou 730000,  China}
\affiliation{School of Nuclear Science and Technology, University of Chinese Academy of Sciences, Beijing 100049, China}
\author{X.~L.~Tu}
\affiliation{Key Laboratory of High Precision Nuclear Spectroscopy and Center for Nuclear Matter Science, Institute of Modern Physics, Chinese Academy of Sciences, Lanzhou 730000,  China}
\author{Y.~M.~Xing}
\affiliation{Key Laboratory of High Precision Nuclear Spectroscopy and Center for Nuclear Matter Science, Institute of Modern Physics, Chinese Academy of Sciences, Lanzhou 730000, China}
\author{J.~C.~Yang}
\affiliation{Key Laboratory of High Precision Nuclear Spectroscopy and Center for Nuclear Matter Science, Institute of Modern Physics, Chinese Academy of Sciences, Lanzhou 730000,  China}
\author{Y.~J.~Yuan}
\affiliation{Key Laboratory of High Precision Nuclear Spectroscopy and Center for Nuclear Matter Science, Institute of Modern Physics, Chinese Academy of Sciences, Lanzhou 730000,  China}
\author{Q.~Zeng}
\affiliation{East China University of Technology, Nanchang 100049, People's Republic of China}
\author{X.~Zhou}
\affiliation{Key Laboratory of High Precision Nuclear Spectroscopy and Center for Nuclear Matter Science, Institute of Modern Physics, Chinese Academy of Sciences, Lanzhou 730000,  China}
\affiliation{School of Nuclear Science and Technology, University of Chinese Academy of Sciences, Beijing 100049, China}
\author{X.~H.~Zhou} 
\affiliation{Key Laboratory of High Precision Nuclear Spectroscopy and Center for Nuclear Matter Science, Institute of Modern Physics, Chinese Academy of Sciences, Lanzhou 730000,  China}
\author{W.~L.~Zhan}
\affiliation{Key Laboratory of High Precision Nuclear Spectroscopy and Center for Nuclear Matter Science, Institute of Modern Physics, Chinese Academy of Sciences, Lanzhou 730000,  China}
\author{S.~Litvinov}
\affiliation{GSI Helmholtzzentrum f{\"u}r Schwerionenforschung, Planckstra{\ss}e 1, Darmstadt, 64291 Germany}
\author{T.~Uesaka}
\affiliation{RIKEN Nishina Center, RIKEN, Saitama 351-0198, Japan}
\author{Y.~Yamaguchi}
\affiliation{RIKEN Nishina Center, RIKEN, Saitama 351-0198, Japan}
\author{T.~Yamaguchi}
\affiliation{Department of Physics, Saitama University, Saitama 338-8570, Japan}
\author{A.~Ozawa}
\affiliation{Insititute of Physics, University of Tsukuba, Ibaraki 305-8571, Japan}
\author{B.~H.~Sun}
\affiliation{School of Physics and Nuclear Energy Engineering, Beihang University, Beijing 100191,  China}

\date{\today}

\begin{abstract}

Isochronous Mass Spectrometry (IMS) in heavy-ion storage rings is an excellent experimental method for precision mass measurements of exotic nuclei. In the IMS, the storage ring is tuned in a special isochronous ion-optical mode. Thus, the mass-over-charge ratios of the stored ions are directly reflected by their respective revolution times in first order. However, the inevitable momentum spread of secondary ions increases the peak widths in the measured spectra and consequently limits the achieved mass precision.  In order to achieve a higher mass resolving power, the ring aperture was reduced to 60 mm by applying a mechanical slit system at the dispersive straight section. The momentum acceptance was reduced as well as better isochronous conditions were achieved. The results showed a significant improvement of the mass resolving power reaching $5.2 \times 10^{5}$, though at the cost of about 40\% ion loss.

\end{abstract}
\maketitle
\section{Introduction}
The nuclear mass is one of fundamental properties of a nucleus. 
It provides valuable information on the nuclear binding energy that embodies the summed effect of all interactions among ingredient nucleons. 
Accurate mass values play an essential role in various research subjects ranging from chemistry to stringent tests of weak interaction and the Standard Model. 
In general, the required accuracy depends on the research subject being explored \cite{Blaum}. 
For example, investigations of the evolution of shell closures requires relative mass precision better than $\delta m / m ~=~10^{-6}$, 
while the test of the conserved vector current (CVC) hypothesis requires a relative precision better than $10^{-8}$. 
Most of the nuclei with unknown masses are far from the valley of $\beta$-stability. 
Hence, their precision mass measurements are constrained by their low production cross-sections and short half-lives. 
Isochronous mass spectrometry (IMS) established in storage rings has been proven to be a powerful tool 
for mass measurements of exotic nuclei with short half-lives even down to several tens of microseconds~\cite{zq,bhsun1}.
Furthermore, IMS allows for sensitive detection of a single ion by using a time-of-flight (ToF) detector \cite{trot,detector}. 
In recent years, many important questions concerning nuclear structure and nuclear astrophysics 
have been addressed by applying IMS at the Experimental Storage Ring (ESR) at GSI, Darmstadt, Germany \cite{JSt,PRep,GSI,YuriReview,RKn2} 
and the Experimental Cooler-Storage Ring (CSRe) at IMP, Lanzhou, China \cite{tuxiaolin,tuprl,zyh,IMP,Yan13,peng14,xx,ZhangReview,zpplb,xing18,fu18,zpprc,smz}. 
Moreover, several new storage ring facilities aiming at the IMS mass measurements are being constructed or planned \cite{ZhangReview,Walker}.
\begin{figure}\centering
	\includegraphics[angle=0,width=9 cm]{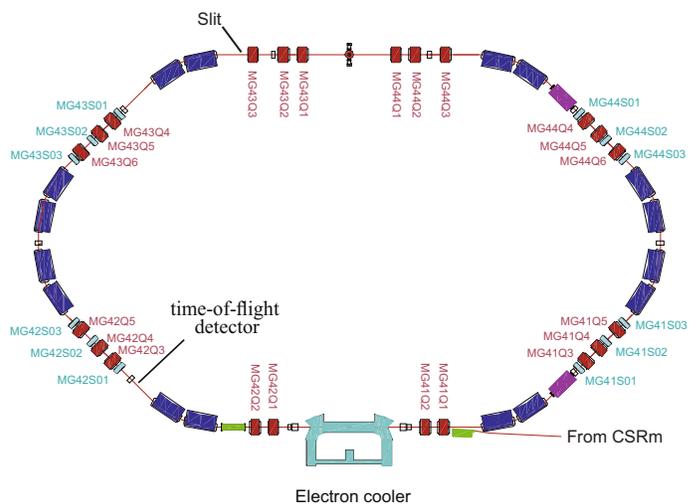}
	\caption{(Colour online) The layout of the CSRe. The positions of the ToF detector and the introduced mechanical slit are indicated.
		\label{Fig1}}
\end{figure}

For particles stored in a ring, their revolution times ($T$) depend on their mass-to-charge ratios ($m/q$) and momenta ($p$).  
In a first-order approximation one can write \cite{Wollnik,Hausmann,Hausmann2,Franzke}:
\begin{align}
\frac{\Delta T}{T}=\frac{1}{\gamma_t^2} \frac{\Delta(m/q)}{m/q}+( \frac{1}{\gamma_t^2}-\frac{1}{\gamma^2})\frac{\Delta p}{p},\label{eq1}
\end{align}
where ${\gamma}$ is the relativistic Lorentz factor of the stored particles and ${\gamma_t}$ denotes the transition energy of the storage ring, 
an ion-optical parameter determined by the ring lattice. 
Various efforts are made to minimize the second term on the right hand side of the above equation \cite{Franzke}, which directly affects the achievable mass resolving power. 
In IMS, the storage ring is tuned into the isochronous ion-optical mode, characterized by a certain ${\gamma_t}$ value.
The ions are injected with energies corresponding to ${\gamma}\approx\gamma_t$.
As a consequence, the revolution times for these species are directly determined by their $m/q$ and are (in first order) independent of their momentum spreads. 
It is clear that due to a limited magnetic rigidity ($B\rho=mv\gamma/q$) acceptance of the ring, this condition is fulfilled for a limited range of $m/q$ values. 
The narrow $m/q$ region in which the isochronous condition ${\gamma_t} \approx\ {\gamma}$ is roughly fulfilled is called the isochronous window \cite{window}. 
Typically, the absolute value of phase-slip factor ${\eta}$, defined as $1/{\gamma_t^2}-1/{\gamma^2}$, is as small as $10^{-3}$ in the isochronous window
and increases rapidly depending on the proximity of ${\gamma}$ to ${\gamma_t}$. 
The above considerations assume that ${\gamma_t}$ is constant over the entire acceptance of the ring.
In practice, due to the field imperfections and the chromatic aberrations  of magnetic elements, the parameter ${\gamma_t}$ 
has a dependence on the closed orbit (magnetic rigidity). 
For more details, the reader is referred to Fig. 4.19 in Ref. \cite{thesis} and Fig. 3 in Ref. \cite{Dolin} 
where ${\gamma_t}$ as a function of $B\rho$ is illustrated for the case of the ESR. 
There are investigations on how to minimize such nonlinearities by introducing higher multipole magnetic fields like octupoles or even decupoles \cite{Sergey}.
Thus, the large momentum spread due to the nuclear reaction process and the non-constant ${\gamma_t}$ contribute to the spread of the measured revolution times
and limit the mass resolving power. 
To achieve a higher mass resolution, a pioneer technique, called \emph{B}$\rho$-tagging, was realized at GSI \cite{10,11,12}. 
There, a narrow slit system was utilized at the second dispersive focal plane of the in-flight fragment separator FRS to define (restrict) the \emph{B}$\rho$ spread 
of transmitted and injected fragments to $\Delta(B\rho)/(B\rho)=1.5 \times 10^{-4}$, while the injection acceptance of the ESR is $\Delta(B\rho)/(B\rho)\approx10^{-3}$. 
As a result, mass resolving power of about $5 \times 10^{5}$ was achieved though at a cost of dramatically reduced statistics.

\begin{figure}[h]
	\includegraphics[angle=0,width=9 cm]{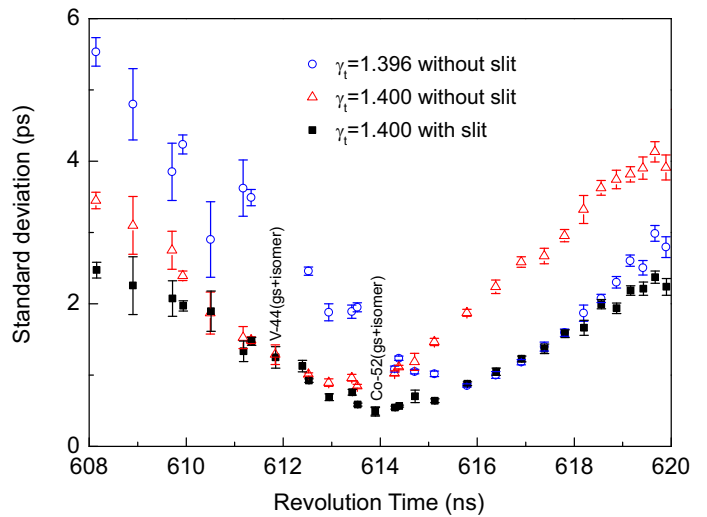}
	\caption{(Colour online) Standard deviations of the measured revolution time peaks, $\sigma(T)$. 
	Data from three experimental settings after correction for the effect of magnetic field instabilities are shown. 
	Open circles and open triangles represent results from the settings without using the slit with ${\gamma}_{t}$ = 1.396, 1.400, respectively. 
	Solid squares show results from ${\gamma}_{t}$ = 1.400 setting using the silt system.
		\label{Fig2}}
\end{figure}

To restrict  simultaneously the momentum spread and the parameter ${\eta}$, a metallic slit was installed in the storage ring CSRe. 
Fig. \ref{Fig1} illustrates the schematic view of the CSRe, in which the positions of the slit and the ToF detector are also shown.
This technique has been utilized in the experiment aiming at mass measurements of $^{58}$Ni projectile fragments. 
By application of the in-ring slit, the mass resolving power about $5.2 \times 10^{5}$ (sigma value) has been achieved, 
to be compared to $1.8 \times 10^{5}$ \cite{tuxiaolin,zyh} in previous experiments without using the slit.
As a highlighted result, low-lying isomers in $^{44}$V and $^{52}$Co are well resolved from the corresponding ground states. 
The mass values and their interpretation have been discussed in Refs. \cite{xx,zpplb,zpprc}. 
In this contribution, unpublished details of the experiment and data analysis are presented.

\section{Experiment and Results}
The experiment was conducted at the Heavy Ion Research Facility in Lanzhou (HIRFL) and Cooler Storage Ring (CSR) accelerator complex. 
The high-energy part of the facility consists of the heavy ion synchrotron CSRm, 
the experimental ring CSRe coupled to CSRm by an in-flight fragment separator RIBLL2 \cite{Xia jiawen}. 
The short-lived nuclei of interest were produced in projectile fragmentation reaction of $^{58}$Ni$^{19+}$ primary beams 
at a relativistic energy on a beryllium-9 target with a thickness of 15 mm. 
At these energies, the produced fragments are fully ionized.
The fragments were selected  by RIBLL2 within a certain \emph{B}${\rho}$ acceptance.
A cocktail beam was injected into the CSRe.
In our context, the CSRe has a relatively large \emph{B}${\rho}$ injection acceptance. 
The transition energy of CSRe was set to ${\gamma}_{t}$ = 1.400 in the isochronous ion-optical mode. 
In order to set the best isochronous condition for $^{52}$Co$^{27+}$, which is the primary goal in this experiment, 
the ring was set to a fixed magnetic rigidity of \emph{B}${\rho}(^{52}$Co$^{27+})$ = 5.8474 Tm, 
calculated for $\gamma$ = ${\gamma}_{t}$ = 1.400. 
Also the magnetic rigidity of the beam-line RIBLL2 was set to this value to allow for an optimal transmission. 
The energy of the primary beam $^{58}$Ni$^{19+}$ was selected to be 467.91 MeV/u according to the calculation via LISE++  
program \cite{lise} so that the $^{52}$Co$^{27+}$ ions had the most probable velocity with $\gamma$ = 1.400 after the exit from the beryllium target.

\begin{figure}[h]\centering
	\includegraphics[angle=0,width=8.5cm]{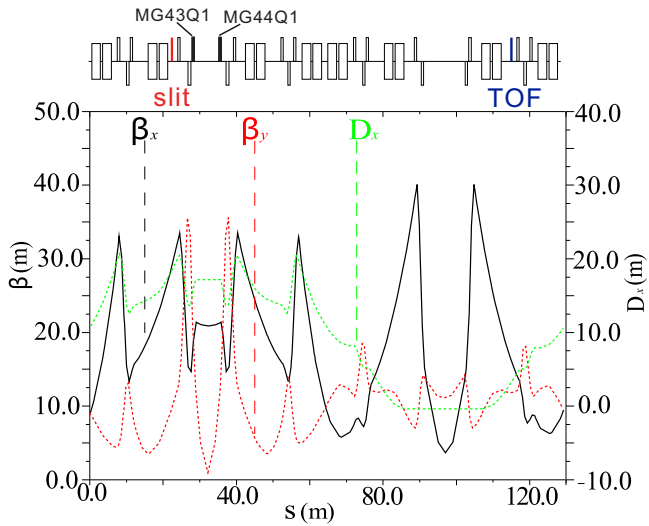}
	\caption{(Colour online) The ${\beta}$-functions and the dispersion function of the CSRe as a function of the orbit length. 
		The thick black line, red dotted line and green dashed line represent the ${\beta}$$_{x}$, ${\beta}$$_{y}$ and \emph{D}$_{x}$ functions, respectively. 
		The positions where the slit and the TOF detector are installed are indicated on the top with a red rectangle and blue rectangle, respectively.
		\label{dispersion}}
\end{figure}

A ToF detector was used to measure the revolution times of the stored ions. 
The detector is based on the detection of secondary electrons released from the surface of a carbon foil installed inside the ring aperture \cite{detector}.
The stored ions penetrate the foil at each revolution. 
Ideally, the electrons are released at each passage of an ion through the detector.  
The electrons were guided to a set of micro-channel plates (MCPs) by perpendicularly arranged electrostatic and magnetic fields. 
The timing signals were recorded directly by a high-performance digital oscilloscope Tektronix DPO 71254. 
For each injection the recording (sampling) time was set to 300 $\mu$s, to be compared to 200 $\mu$s in previous experiments~\cite{tuprl,zyh,window}. 
The revolution times of each ion was extracted from the timing signals. 
After correction for the time drifts due to instabilities of magnetic fields, 
masses of  nuclides of interest were obtained from the final revolution time spectrum. 
Our conventional procedure for the data analysis has been described in detail in Refs. \cite{tuxiaolin, zpprc}.

After 12-hours data accumulation (see blue open circles in the Fig. \ref{Fig2}), 
it has been found that nuclides with revolution times around 616 ns had the minimum standard deviation $\sigma(T)$, 
while the revolution time of $^{52}$Co$^{27+}$ ions was about 614 ns. 
Since this is a good indicator for the isochronous condition, the ring was obviously not optimized for $^{52}$Co$^{27+}$. 
According to the measured experimental data, the transition energy ${\gamma}_{t}$ of the CSRe was about 1.396. 
This slight deviation of  ${\gamma}_{t}$ from the aimed value was mainly caused by the imperfections of the ring magnetic fields.  
A first order optimization of the ion-optical isochronous setting was made via modifications of the quadrupole magnetic field strengths~\cite{gx}.  
The current of a family of quadrupoles, MG43Q1 and MG44Q1, was increased by 0.4$\%$. 
In this way, the ${\gamma}_{t}$  of the CSRe was corrected to 1.400. The success of this optimization was confirmed after a 8-hours data accumulation (see red triangles in the Fig. \ref{Fig2}), where the nuclides with revolution times around 613 ns showed the minimum standard deviation $\sigma(T)$.

\begin{figure}[h]\centering
	\includegraphics[angle=0,width=9 cm]{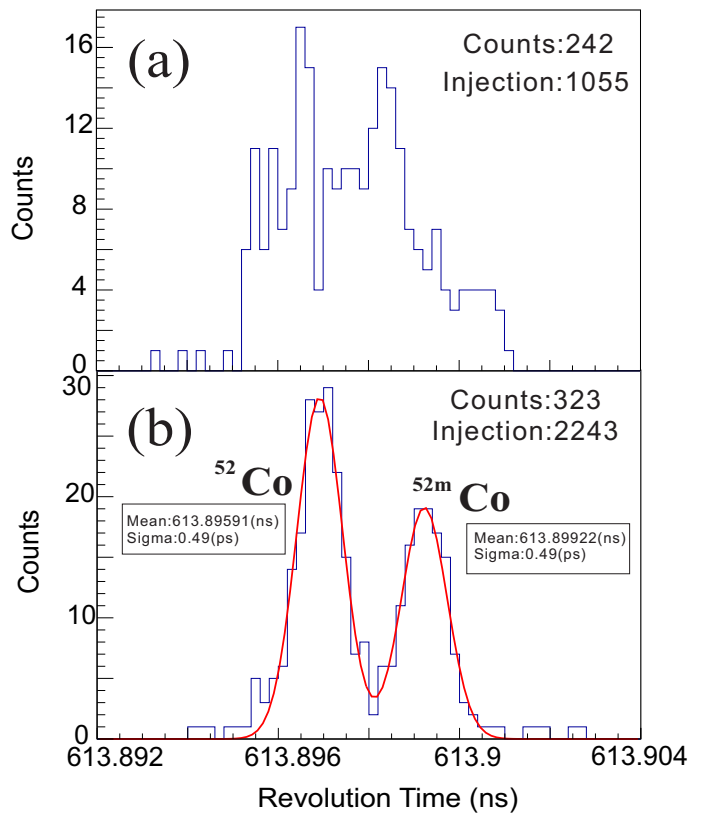}
	\caption{(Colour online) The revolution time spectra of $^{52}$Co zoomed in a time window of 613.892 ns $\leq$ \emph{t}  $\leq$ 613.904 ns. The top and bottom panels show the spectra before and after the application of the slit, respectively. The displayed results in the latter panel are from double-Gaussian chi-squares fitting.
		\label{Fig4}}
\end{figure}

The excitation energy of the low-lying isomeric state in $^{52}$Co is about 390 keV, 
which is inferred from its mirror nucleus $^{52}$Mn~\cite{52Mn} 
regardless of isospin-symmetry breaking.
According to Eq. \ref{eq1}, the corresponding difference of revolution times between the isomeric and ground states in $^{52}$Co is about 3 ps. 
However, from the results of the described two settings, the minimum standard deviation is about 1 ps for the nuclides with the best isochronous condition. 
To achieve better resolution of the isomer from the ground state in the revolution time spectrum, a higher mass resolving power is needed.

For this purpose, a mechanical slit limiting the ring aperture has been installed.
In principle, it should be installed at the place where the dispersion is large.
Its  actual position was determined according to the simulation for  $\beta$-functions and dispersion function of the ion-optical setting of the CSRe as shown in Fig.~\ref{dispersion}. 
The dispersion at the position of the slit was estimated to be 20 cm/\%. 
The width of the slit set to be was 60 mm, corresponding to the momentum acceptance of the CSRe of $\Delta p/p$ $\sim~\pm$ 0.15$\%$, 
while this value was $\sim~\pm$ 0.2$\%$ in the previous experiments under the same optical settings but without the application of the slit~\cite{zyh}.
 
This method effectively improves the precision of the revolution time measurement in comparison with the other settings as demonstrated in Fig. \ref{Fig2}. 
The obtained standard deviations of the revolution times are shown as black squares in this figure. 
There were two changes that should be addressed. The first one was that the revolution time, 
of which nuclide have the minimum $\sigma(T)$, shift from 613 ns to 614 ns even though the current of any magnet was not change at all.
As discussed in the introduction, this shift is mainly due to the none constant value of $\gamma_t$ in the full momentum acceptance.  The $\gamma_t$ value discussed here was an average result. By using the slit, the $\gamma_t$ parameter was restricted in a smaller momentum acceptance, and thus the average value of $\gamma_t$ changed.  
The second point is that the  resolving power was improved by about a factor of two for $^{52}$Co. 
Fig.~\ref{Fig4} clearly illustrates that the low-lying isomer in $^{52}$Co with $E_x$ of about 390 keV was now well resolved from the ground state. 
Furthermore, both the nuclides of interest and nuclides used as references in the mass calibration procedure benefited from this method. 
As a result, the statistical error and the fitting error in the mass calibration could be reduced leading 
to an unprecedented mass precision of 5 $\sim$ 10 keV reached so far in the IMS for short-lived nuclei.

\begin{figure}[htbp]
	\includegraphics[angle=0,width=9 cm]{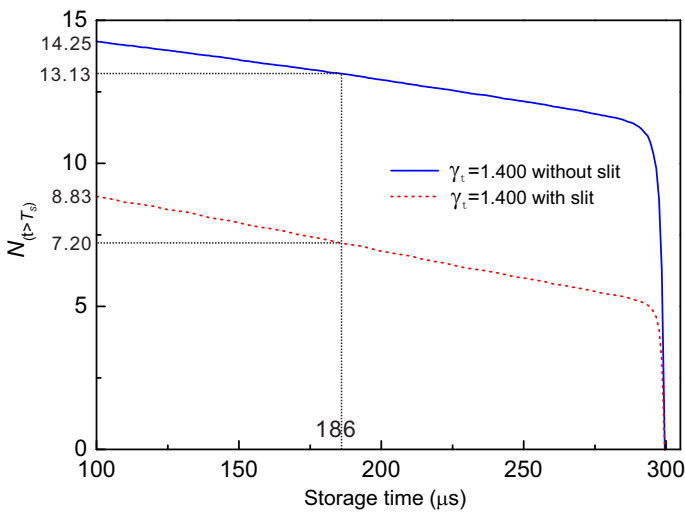}
	\caption{(Colour online) Beam loss as a function of the storage time ranging from 100 $\mu$s to 300 $\mu$s. All ions  with revolution time in the window
	from 608 ns to 620 ns were counted. $N_{(t>T_s)}$ is the average number (per injection) of the  ions whose storage time $t$ longer than a given $T_s$.
	The blue solid-line and red dash-line represent the results from the setting of ${\gamma}_{t}$ = 1.400 before and after employing the slit, respectively. 
	\label{lifetime}}
\end{figure}

 Similar to the \emph{B}$\rho$-tagging method applied at the FRS-ESR,  the negative consequence of the application of the slit is that the available orbitals in the CSRe were reduced leading to the loss of valuable particles due to the smaller acceptance. We compared the average number of survived ions before and after the application of the slit.
As shown in Fig. \ref{lifetime},  a great reduction of about 40\% was clearly seen after using the slit. Furthermore, the continuous decease of the average number in both cases reveals that the beam gradually lose when circling in the ring. In the previous experiments, only those ions which circulated for more than about 300 revolutions (186 $\mu$s) in CSRe were considered in the data analysis~\cite{tuxiaolin}. Obviously, some ions do not survive for so long time. Before using the slit, the average number decreased by 8\% (from 14.25 to 13.13) when compared ions with storage time longer than 100 $\mu$s and 186 $\mu$s, while decreased by 18\% (from 8.83 to 7.20) after using the slit. The loss of ions corresponding to the storage time was amplified by using the slit. 

The uncertainty of the revolution time of each stored ion, that was extracted from periodic timing signals, 
contains two contributions, the finite emittance (defined as a deviation from the reference particle) of the ion and the time resolution of the ToF detector.  
The influence of the former can be eliminated by averaging the data from a large number of revolutions. 
The uncertainty contribution from the latter has been estimated in Ref.~\cite{zq} to be
\begin{align}
{\delta T} {\approx}\frac{3.64\sigma}{\sqrt{\varepsilon M^3}},\label{eq2}
\end{align}
where $\varepsilon$ is the detection efficiency, $\sigma$ (about 50 ps) is the time resolution of the ToF detector, and $M$ is the number of turns that ions were stored in the ring. 
$\varepsilon$ varies from 20\% to 90\% depending on the total number of timing signals in the detector in one injection and on the proton number of the ion~\cite{detector,zhangwei}. 
In this experiment, it is about 50\% for ions with proton number around 20. 
For an ion that was stored for more than 100 $\mu$s, namely for more than 150 turns, the uncertainty of the revolution time was better than 0.1~ps. 
This value shall be compared to the standard deviations of the revolution time peaks in the final spectrum of larger than 0.5 ps.

Finally, all ions that circulated for more than 100 $\mu$s were used, leading to an increase of statistics by about 20\%. 
Our present results show that the limitation set on the storage times in the data analysis of the previous experiments was too conservative.

 \begin{figure}[htbp]
 	\includegraphics[angle=0,width=9 cm]{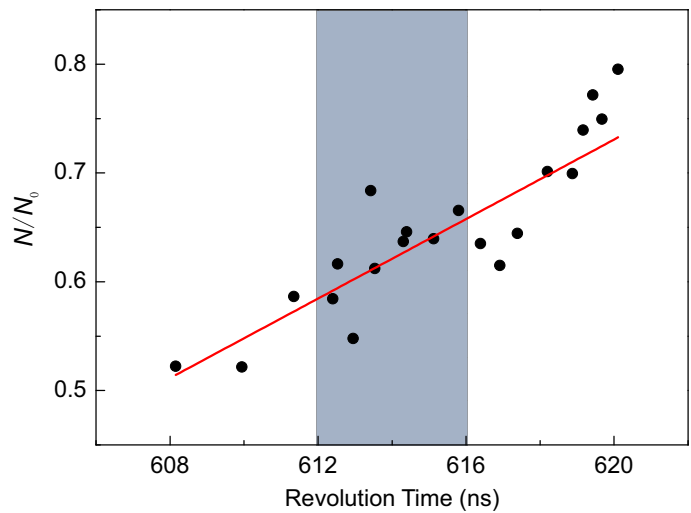}
 	\caption{(Colour online) Ratio of the numbers of stored ions in two settings for different nuclides. \emph{N} and \emph{N}$_{0}$ are the average numbers of ions (per injection) for a given nuclide in the setting of ${\gamma}_{t}$ = 1.400 but with and without the slit, respectively.  The red line in the figure has no meaning and  just guide the eyes.
 		\label{lost}}
 \end{figure}
 
The loss of ions for different species of nuclide, caused by the application of the slit,  was further investigated as shown in Fig. \ref{lost}.
The normalized number for each nuclide is all smaller after using the slit.  
About 30\%-45\% of ions were lost in the revolution-time window from 612 ns to 616 ns, 
where the nuclei of interest were mainly located (indicated by the grey-shadowed region in Fig. \ref{lost}). 

Meaningfully, we found that the \emph{N}/\emph{N}$_{0}$ has a positive correlation with the revolution time. 
A possible explanation for this dependence is that the momentum distribution for each nuclide is different, as discussed in Ref.~\cite{momentum,mom2}. Without the slit, nuclei within the entire acceptance of CSRe can be stored in the ring. After applying the slit, the momentum distributions are restricted in a smaller range. In this experiment, the nuclei with longer revolution times are in general heavier and closer to the projectile, thus have narrower momentum distribution compared with the nuclei with shorter revolution times. Thus, the \emph{N}/\emph{N}$_{0}$ is larger for nuclei with longer revolution time.
To testify this hypothesis, we plan to measure the actual momentum distributions of nuclei in the ring 
by using a double-TOF detector system~\cite{window,sp_nimb} in the forthcoming experiments.

\section{Summary and perspective}

We presented some details of the experiment and data analysis of isochronous mass measurements at the CSRe with the application of an in-ring slit system. 
Owing to the slit, the momentum distribution was reduced and a better isochronicity has been achieved. 
The results have shown that the IMS with an applied slit can lead to a significantly improved mass resolving power. 
In this experiment we achieved $m/\Delta m = 5.2 \times 10^{5}$ (sigma value).  However, the application of the slit leads to the loss of about 40\% of ions. 
Our method was further successfully applied in experiments addressing $^{112}$Sn projectile fragments \cite{xing18}, 
where a slit with a narrower, 50 mm opening, has been introduced resulting in even higher mass resolving power. 

In the past few years, the IMS at the CSRe has been extended by the installation of a double-TOF detector system. 
Several pilot experiments have been done~\cite{xymtof}. 
With the new set-up, the velocity of each stored ion can be measured  in addition to its revolution time.
Thus the $\gamma_t$ as a function of the orbit length could accurately be measured~\cite{crj,wwge}. 
With the latter developments, the mass resolving power of the IMS will likely be further improved without losing statistics.

\section{ACKNOWLEDGMENTS}
The authors thank the staffs in the accelerator division of IMP for providing stable beam. 
This work is supported in part by the National Key  R\&D Program of China (Grant No. 2018YF
A0404401 and No. 2016YFA0400504),
the National Nature Science Foundation of China (Grants No. 11605249, 11605248, 11605252, 11505267, 11575112, and 11575007), 
the CAS External Cooperation Program (Grant No. GJHZ1305), 
the CAS through the Key Research Program of Frontier Sciences (Grant No. QYZDJ-SSW-SLH005), 
the Key Research Program of the Chinese Academy of Sciences (No. XDPB09),
the Helmholtz-CAS Joint Research Group (HCJRG-108),
and by the European Research Council (ERC) under the European Union's Horizon 2020 research and innovation programme (Grant No 682841 ``ASTRUm'').
Y.A.L. acknowledges the support by the CAS President's International Fellowship Initiative Grant (2016VMA
043).
K.B. acknowledges support by the Nuclear Astrophysics Virtual Institute (NAVI) of the Helmholtz Association. 
X.X. thanks the support from CAS "Light of West China" Program.

\end{document}